# Topography preserved microwave plasma etching for top-down layer engineering in MoS$_2$ and other van der Waals materials

Abin Varghese, Chithra H. Sharma and, Madhu Thalakulam*

A generic and universal layer engineering strategy for van der Waals (vW) materials, scalable and compatible with the current semiconductor technology is of paramout importance in realizing all-two-dimensional logic circuits and move beyond the silicon scaling limit. In this letter, we demonstrate a scalable and highly controllable microwave plasma based layer engineering strategy for MoS$_2$ and other vW materials. Using this technique we etch MoS$_2$ flakes layer-by-layer starting from arbitrary thickness and area down to the mono- or the few-layer limit. From Raman spectroscopy, atomic force microscopy, photoluminescence spectroscopy, scanning electron microscopy and transmission electron microscopy, we confirm that the structural and morphological properties of the material have not been compromised. The process preserves the pre-etch layer topography and yields a smooth and pristine-like surface. We explore the electrical properties utilising a field effect transistor geometry and find that the mobility values of our samples are comparable to those of the pristine ones. The layer removal does not involve any reactive gasses or chemical reactions and relies on breaking the weak inter-layer vW interaction making it a generic technique for a wide spectrum of layered materials and heterostructures. We demonstrate the wide applicability of the technique by extending it to other systems such as Graphene, h-BN and WSe$_2$. In addition, using the microwave plasma in combination with standard lithography, we illustrate a lateral patterning scheme for device fabrication.

All-two-dimensional devices consisting of semiconducting, metallic and insulating van der Waals (vW) materials offer a comprehensive-solution to overcome the current scaling limit of the micro-electronic industry. Successful demonstrations of heterostructures by stalking various two-dimensional (2D) systems[1–3] and realization of MoS$_2$ and WSe$_2$ transistors with graphene electrodes and h-BN gate-insulator have imparted a new momentum in this direction.[4–6] The natural abundance, excellent mechanical and chemical stability, and superior electrical characteristics make MoS$_2$ the most sought-after 2D semiconductor for microelectronics technology.[7,8] Transistors with large on/off ratio and high carrier mobility,[9] high-frequency operation,[10] logic circuits such as NOT and NOR,[11] NAND, random access memory and ring oscillators[12] have been demonstrated in mono- or few-layer MoS$_2$. Mono- and few-layer MoS$_2$ crystals are mostly obtained by CVD,[13–16] PECVD,[17] and also by mechanical exfoliation of bulk crystals.[9,18] CVD grown materials often exhibit a large number of domains. Mechanical exfoliation has yielded samples with superior electrical behaviour compared to other methods, lack control on the device thickness and area. The technology aspirations of MoS$_2$ and other vW materials depend on a feasible method to engineer large-area mono- or few-layer materials on-demand.

*School of Physics, Indian Institute of Science Education and Research Thiruvananthapuram, Thiruvananthapuram-695016, Kerala, India.*
*madhu@iisertvm.ac.in*

Though various etching schemes have been explored,[19–29] a scalable and controllable thickness reduction down to the mono- or few-layer regimes starting from arbitrary thickness and area has not been demonstrated. Methods utilizing material chemistry exploit either harsh chemicals[19] or reactive plasma environment, are material specific.[21–23,28] Physical etching involves high temperature annealing,[25,27] laser assisted thermal ablation[20] or plasma environments,[24,26,29] results in degraded electrical properties. Post-etch electrical characterization has not been performed in most cases, except for a few.[19,20,27] A generic top-down approach applicable to any vW material, scalable and compatible with the current fabrication technology, without compromising the electrical properties, would stride towards the realization of all 2D micrelectronics.

In this letter, we demonstrate a controllable layer-by-layer thickness reduction of $MoS_2$ using a forming-gas (FG) microwave plasma. Raman spectroscopy and transmission electron microscopy (TEM) are used for structural characterizations. Atomic force microscopy (AFM), optical-contrast method, stylus profilometry and scanning electron microscopy (SEM) are used for morphological analysis. We perform low-noise transport studies on post-etched devices with back-gated field-effect transistor (FET) geometry, in high-vacuum light-tight environment to explore the electrical properties. Our studies show that the etching process did not degrade the structural and electrical properties from those of the pristine material. Additionally, we demonstrate that our plasma etching strategy can be extended for device patterning using a lithography defined metal mask. We also demonstrate the generic nature of the technique by controllably etching other vW material systems; graphene, $WSe_2$ and h-BN.

The microwave plasma etching system consists of a 2.45 GHz microwave oven[30,31] equipped with a fused-quartz vacuum chamber. We use FG (90% Ar + 10% $H_2$, premixed) plasma for etching all the samples presented in this report unless specified otherwise. The plasma power can be continuously varied by controlling the input voltage to the primary of the high-voltage transformer driving the magnetron.[30] The FG pressure is optimized by fine-control needle valves and gas flow meters. Thermocouple temperature sensors are used to monitor the substrate temperature during the operation.

$MoS_2$ samples are mechanically exfoliated from bulk crystals onto a clean Si wafer hosting a ~300 nm thick thermal oxide layer. Prior to the plasma treatment, the exfoliated samples are examined for surface smoothness, area and number of plateaus (regions with the same number of layers) using a trinocular optical microscope (Olympus BX51M). The colour change, resulting from the thickness reduction of $MoS_2$ due to an increase in transparency on approaching the few-layer-limit, is used as a primary estimate of thickness (see Supporting Information, S1). An accurate measure of thickness is obtained using AFM, stylus profilometry and, a well calibrated optical contrast method.[32] As opposed to AFM, profilometry and optical contrast methods provide time-efficient thickness determination avoiding the exposure of the sample to the atmosphere for a longer duration.

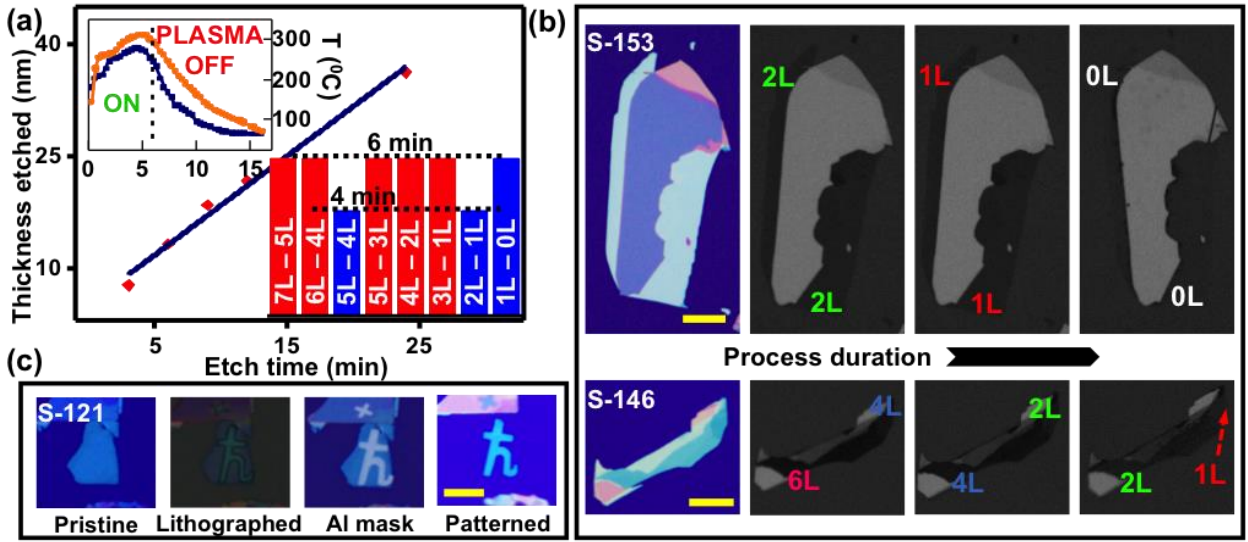

Figure 1: (a) Plot of plasma exposure duration v/s thickness reduction for the coarse etch-rate. The bar graph in the bottom inset denotes the etching time for one and two layers using the fine etching technique. The plot in the top inset is the temperature of substrate v/s process duration for coarse (orange) and fine (blue) etching techniques. (b) First column shows optical images of pristine exfoliated $MoS_2$ samples S-153 and S-146 respectively and the subsequent columns show R-channel contrast images of a series of controlled layer reduction steps on both the samples. Corresponding thickness at each stage is denoted in terms of layer numbers followed by L. The cut on S-153 (top-right) marked by the red-box is caused by an unintentional exposure of high power laser. (c) Step-by-step demonstration of the device-patterning technique. All scale bars correspond to 20 μm.

We devise two etching recipes, the coarse and the fine, characterized by 25% input power to the magnetron with 0.1 mbar of FG and 10% input power to the magnetron with 0.05 mbar of FG respectively. The coarse etch provides a thickness reduction of ~1.5 nm/minute as shown in Figure 1 (a). The etch rate is extracted by determining the thickness of the samples using a stylus profilometer (calibrated against AFM) after successive etching stages. Each point in the trace represents data taken from five samples. The fine etch is used for the precise control of the thickness below ~ 10 layers. AFM and the optical-contrast method are used for thickness analysis before and after the etch (see Supporting Information, S2). A bar graph depicting the time taken for etching two-layers and one-layer is given in the bottom-right inset. The numbers on each bar represent the initial and the final number of layers and the height corresponds to the etching time. Etching two consecutive layers takes six minutes and one-layer except the monolayer takes 4 minutes. Complete removal of the monolayer takes slightly longer,[27,28] 6 minutes, which provides the process better controllability preventing accidental removal of the whole sample. Each data point has been verified multiple times and we do not find any deviation from this observation. Using a combination of the coarse and the fine etch we are able to thin samples of arbitrary area and thickness down to any desired number of layers; coarse etch is employed down to ~10 layers and fine etch is used for etching further down. We note that the temperature rises to ~310 °C for the coarse (the orange trace) and ~275 °C for the fine (the blue trace) process owing to microwave heating.

Figure 1 (b) shows optical images of two representative samples (S-153 and S-146) having initial dimensions larger than 100 μm laterally and 100 nm in thickness, etched down using a combination of the coarse and the fine etch. The images, arranged horizontally, depict the etching chronology; the first column shows bright-field images of the pristine samples and the following columns show the red-channel optical images (extracted using ImageJ) for the last three consecutive etching steps. The regions with six-layer (6L) and four-layer (4L) of the sample S-146 has been thinned down to four-layers (4L) and two-layer (2L) respectively in the successive step, keeping the pre-etch layer profile unaltered. Also, the two-layer (2L) region of sample S-153 has been reduced to monolayer (1L) and etched-off completely (0L) in the successive etching steps. The optical contrast method is accurate only for samples less than eight layers. In this regime we have verified that the layer reduction is uniform and layer-by-layer, preserving the initial layer-topography across the sample (see Supporting Information, S2).

Our etching process in combination with standard lithography techniques can also be used for patterning samples for device applications. We use Aluminium as the etch-mask. Mechanically exfoliated samples are patterned using standard photo/electron-beam lithography followed by Aluminium metallization and lift-off. The unmasked region is etched off completely; thereafter the Aluminium mask is removed using NaOH solution. Various stages of the process are demonstrated in Figure 1 (c).

We perform structural characterization of the etched-down samples using Raman spectroscopy with 532 nm laser (Horiba XploRA ONE), Figure 2 (a). All the plasma-etched samples, down to the monolayer, exhibit the characteristic in-plane $E^1_{2g}$ and out-of-plane $A_{1g}$ Raman peaks of $MoS_2$, implying the structural integrity has been retained during the process. Plasma treated samples can exhibit Raman shifts different from that of the pristine samples [28] and our plasma-etched monolayers (1L) exhibit a frequency difference of 22 cm$^{-1}$ between the $E^1_{2g}$ and $A_{1g}$ peaks. Disorder and defect can lead to emerging peaks and broadening of the characteristics peaks in the Raman spectra,[24,33] which we do not observe. Also, we do not observe any characteristic $MoO_3$ peaks ruling out any oxidation of $MoS_2$ during the process (see Supporting Information, S3). Figure 2 (b) shows the photoluminescence spectra obtained for the etched down monolayer showing the characteristic direct-gap emission peak of ~1.82 eV. The cut on the sample shown in the red-box of S-153, Figure 1 (b), is caused by an unintentional exposure of high-power laser resulting in thermal ablation of $MoS_2$.[20]

We examine the atomic structure of the post-etch few-layer $MoS_2$ sample using a 120 keV HR-TEM, Fig. 2 (c). The inset is the magnified TEM image showing a highly ordered lattice. The images do not show any significant irregularities, point-defects or distortions in the crystal structure. We extract inter-planar distances of ~ 0.27 nm and ~ 0.15 nm for the (100) and (110) planes respectively. Our TEM data and the lattice parameters are similar to the values reported for pristine sample.[24]

We utilize an AFM in the tapping mode (Bruker Dimensions Edge) to characterize the surface roughness and uniformity of the plasma etched samples. In Figure 2 (d) we focus on the post-etch surface morphology of S-153. The AFM height profile corresponds to the red-channel image given in the second column of Figure 1 (b). Either of the two-layer regions (2L) has been etched down starting from more than

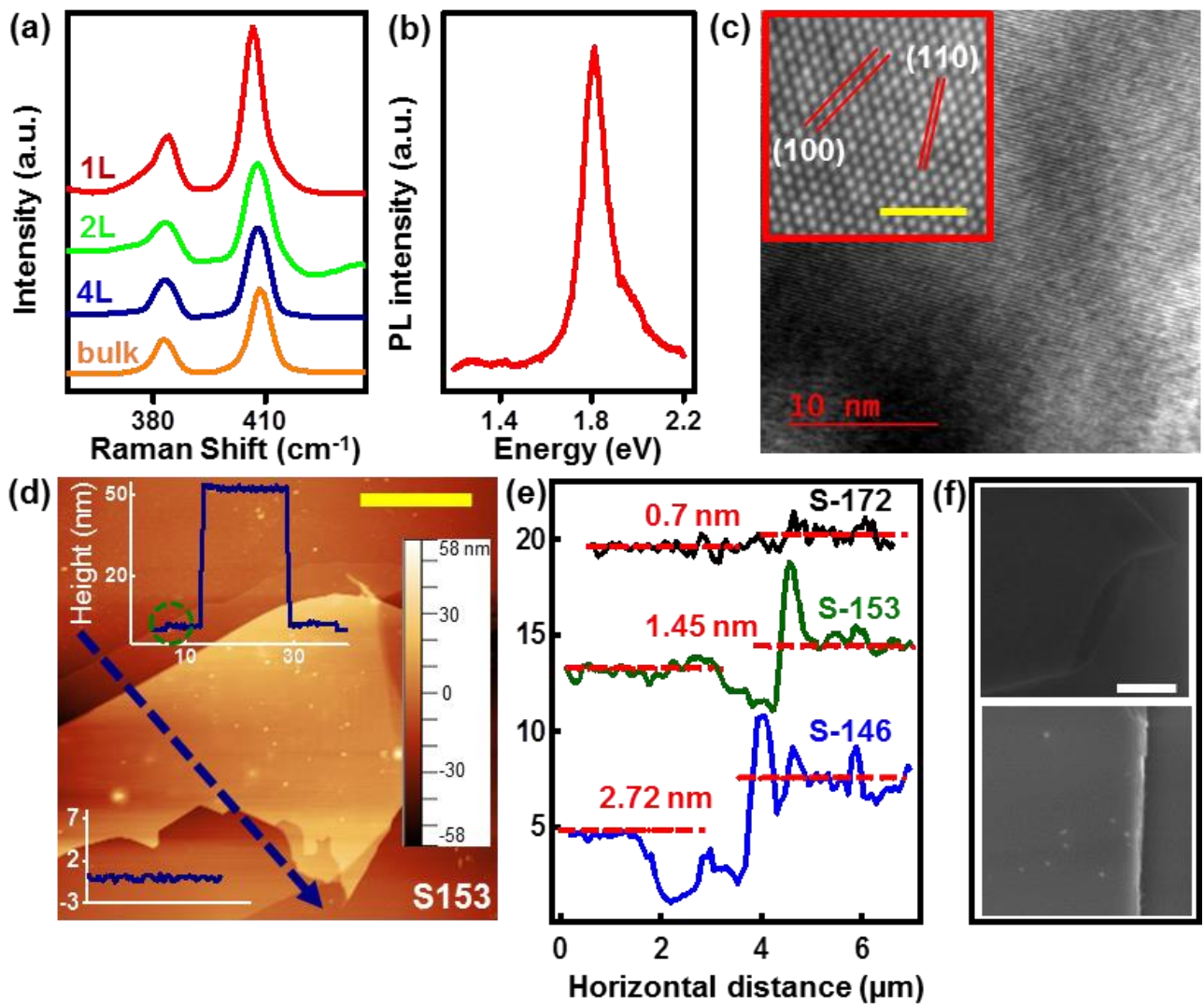

Figure 2: Structure & Morphology: (a) Raman Spectra for monolayer, bilayer, four-layer and bulk plasma etched samples showing characteristic $MoS_2$ Raman peaks. (b) Photoluminescence spectra of etched monolayer $MoS_2$. (c) HR-TEM images of few layer plasma etched $MoS_2$. Inset shows a magnified image showing the atomic structure. The (100) and the (110) planes are marked as shown. Scale bar is 2 nm. (d) AFM height profile image of S-153. The extracted surface profile along the marked blue arrow is shown in the top inset. The bottom inset shows the line profile extracted along the 2L region showing a smooth surface morphology. Scale bar corresponds to 20 μm (e) The extracted AFM height profiles for 1L, 2L and 4L samples obtained after plasma thinning of thicker parent flakes. The bilayer profile is a magnified view of the profile marked by green circle in Figure 2(d). 1L and 4L profiles are extracted from the AFM data of S-172 and S-146 respectively (S. I). (f) SEM images of plasma-etched (top) and pristine (bottom) $MoS_2$ samples. Scale bar is 500 nm.

thirty layers using a combination of the coarse and the fine etch processes. The top inset shows the line-profile taken along the blue dashed arrow showing three distinct plateaus. The pristine exfoliated sample had also featured similar plateaus [Figure 1 (b)]. The etched 2L region exhibits a smooth and homogenous surface morphology as observed from the height profile and the line-profile (bottom inset). We extract the RMS surface roughness (Rq) and the average surface roughness (Ra) of 0.42 nm, and 0.32 nm respectively. Rq and Ra of a few other plasma-etched samples studied for this report are tabulated in S4 of Supporting Information. We note that the Rq and Ra values of our plasma etched samples are comparable to that of the pristine samples.[34,35] The thickness values of 1L (0.7 nm), 2L (1.45 nm), and 4L

(2.72 nm) extracted from the AFM height profiles as shown in Figure 2 (e) agree with that of the pristine samples. The 2L thickness was obtained from the AFM height profile of S-153 shown in Figure 2(d). The 1L and 4L thickness were extracted from the AFM height profiles of S-172 and S-146 respectively (see Supporting Information, S5). Figure 2(f) shows SEM image of the plasma-etched sample (top) which shows a smooth surface morphology similar to the pristine exfoliated sample (bottom).

In contrast to the plasma or temperature assisted layer reduction techniques we observe a residue-free complete removal of $MoS_2$ layers and the post-etch surface is smooth and flat, like pristine samples. Our etching process involves FG microwave plasma accompanied by thermal annealing due to microwave heating. Layer removal due to thermal annealing involves either the sublimation of $MoS_2$ which requires a much higher temperature, or oxidation of $MoS_2$ to $MoO_3$ and subsequent sublimation of $MoO_3$.[25,27] Both of these processes require extensive duration compared to the one described in this manuscript. In addition, thermal annealing in air is accompanied by a lateral area reduction[27] while annealing in Argon atmosphere yields non-uniform post-etch morphology,[25] both we do not observe.

Now we examine the electrical transport properties of plasma-etched samples. The devices exploit a standard back-gated FET geometry. The source and the drain contacts are realized using standard electron-beam or photolithography followed by Cr/Au metallization. All the electrical measurements are performed in high vacuum (~$10^{-6}$ mbar) unlit environment. The field effect mobility values are calculated from the conductance v/s back-gate voltage (transconductance) using the formula $\mu = L/(WC_{ox}) \times (dG/dV_g)$ where, $C_{ox} = \varepsilon_{ox}/d_{ox}$, $\varepsilon_{ox}$ the permittivity of the oxide, $d_{ox}$ the thickness of the oxide, L and W the length and width of the channel respectively and, G the conductance of the device. We present transport data from three samples; (i) S-91, part of the sample has been etched down, (ii) S-136, which has been uniformly thinned down from bulk using the coarse etch process and, (iii) S-172 which has been etched down to 5 layers from ~10 layers with the fine etch process.

For a direct comparison of the electrical transport on the etched and unetched regions on the same device, we focus on Sample S-91, a portion of which has been thinned down to 12 nm from ~48 nm by the coarse etch technique. Figure 3 (a) shows the optical image of the device (top); left-half (pink coloured region) represents the unetched side of thickness ~48 nm and the right-half (blue coloured region) represents the etched side of thickness 12 nm. The Raman spectra (bottom panel of Figure 3 (a)) of both the etched (pink trace) and the unetched (blue trace) sides exhibit the characteristic Raman peaks of $MoS_2$ confirming the structural quality of the sample. The AFM height profile of the sample is given in S6 of Supporting Information. Contacts are made on to both the etched and the unetched sides using electron-beam lithography. Figure 3 (b) shows the conductance as a function of back-gate voltage for the etched (blue trace) and the unetched (pink trace). Inset to the Figure 3(b) shows the 2-probe current-voltage characteristics of the etched (blue trace) and the unetched (pink trace) regions exhibiting a clear ohmic behaviour. We obtain field-effect mobility values of 15.45 $cm^2V^{-1}s^{-1}$ for the etched side and 13.08 $cm^2V^{-1}s^{-1}$ for the unetched side from the transconductance data.

The unetched side is four-times thicker than the etched side. The device area and the ohmic-contact area of both the sides are similar. The 2-probe resistance of the etched side (460 kΩ) is ~14 times that of unetched side (33.8 kΩ). For an exact analysis, one needs to consider the interlayer resistance, the number of layers taking part in the transport and the effective device area of the etched and unetched sides. Apart from being thinner, there are other factors contributing to the larger 2-probe resistance of the etched side; (i) increased scattering as it is close to the thermal-oxide interface, (ii) possible confinement of the charge carriers to the top few layers compared to that of the unetched side due to the plasma induced reduction in the interlayer coupling. The observed reduction in the intensity of $A_{1g}$ peak and the increase in the intensity of $E^1_{2g}$ peak of the etched side with respect to those of the unetched side could imply a reduction in the inter-layer coupling.[36,37]

Next, we discuss the transport studies of sample S-136 which has been etched down to ~20 nm from the bulk regime exploiting the coarse etch. The conductance v/s back-gate voltage is shown in Figure 3 (c). The top-left inset shows the optical image of the device. The post-etch Raman spectra (top-right inset) exhibit the characteristic peaks of $MoS_2$ verifying the structural quality of the sample. The bottom inset shows current-voltage characteristics for various back-gate voltages. We extract a field-effect mobility of ~22.27 $cm^2V^{-1}s^{-1}$ from the transconductance data.

Figure 3 (d) shows the conductance v/s back gate voltage of a sample S-177, subjected to the fine etch process. The sample was etched down to 5-layers starting from a thickness ~10 layers. Top-left inset shows the optical image of the device. The post-etch Raman spectra shown in the top-right inset exhibits the characteristic Raman peaks of $MoS_2$ verifying the structural quality. The current-voltage characteristics for various back-gate voltages are given in the bottom inset of Figure 3 (d). We extract a carrier mobility of 12.27 $cm^2V^{-1}s^{-1}$ from the transconductance data.

The current-voltage characteristics of S-136 & S-177 shown in the bottom insets of Figure 3(c) and Figure 3 (d) respectively exhibit a non-linear behaviour as opposed to that of S-91 in shown in 3 (a). This could have been caused by the chemical residues left by the photolithography process used to fabricate those devices. The threshold voltage is influenced by the sulphur vacancies and it could be used as measure of defect density in the sample.[38] Our plasma etched samples exhibit threshold voltages similar to the pristine ones as we have verified (not shown here) and also reported in the literature.[39] The field effect mobility values for our etched samples are also comparable to the pristine samples with similar thickness.[40]

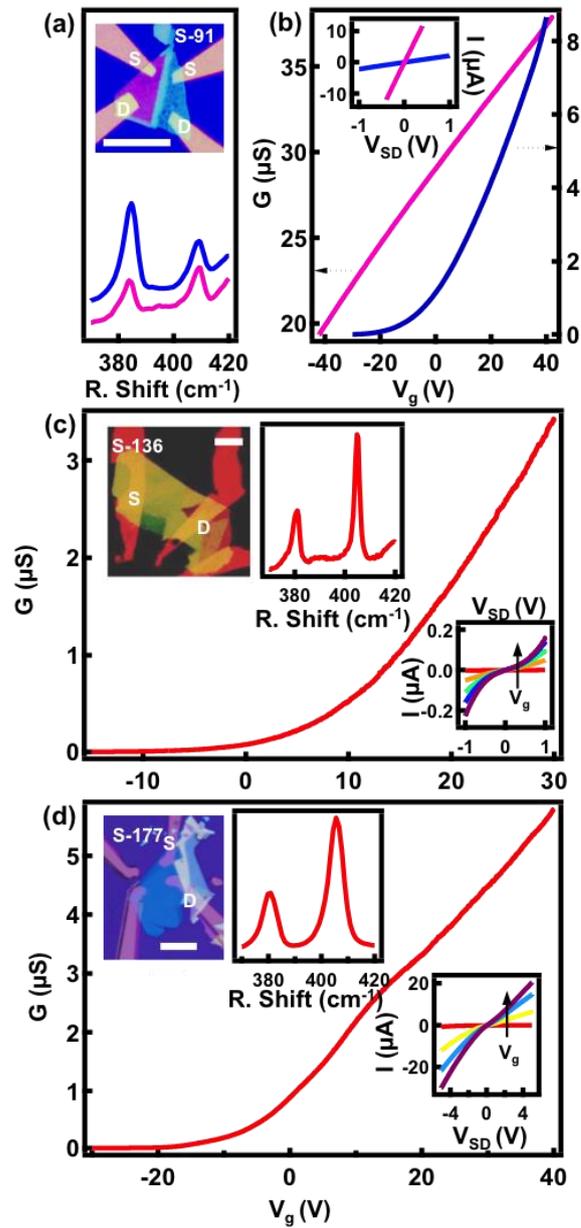

Figure 3: Transport characteristics (a) top: optical image of S-91 with Cr/Au contacts, bottom: Raman spectra of etched (blue) and unetched (pink) sides. (b) Conductance v/s back-gate voltage for etched (blue) and unetched (pink) sides of S-91 with the current-voltage characteristics of etched (blue) and unetched (pink) sides shown in the inset. (c) & (d) Conductance v/s back-gate voltage, optical image of the sample (top-left inset) and post-etch Raman spectra (top-right inset) and current v/s voltage for different back-gate voltages (bottom inset) for S-136 and S-177 respectively. The Source and Drain contacts are marked as S and D respectively. Scale bars are 20 μm.

Next we demonstrate that our plasma etching is a very generic process that can be extended to other van der Waals systems. Figure 4 (a), (b) & (c) show pre- and post-etch optical images of $WSe_2$, h-BN and graphene respectively. Figure 4 (d), (e) & (f) show the respective post-etch Raman spectra obtained from the regions marked by the white circle. The optical contrast shows a uniform thickness reduction across the entire area of the samples. The central region marked by the white circle in Figure 4 (a) has been reduced to 1L as inferred from the Raman spectra.[41] The **D** peak in Figure 4 (f) with a relatively low intensity implies a low defect density in the post-etch graphene which shows that our etching process does not deteriorate the structural quality of the sample.[42]

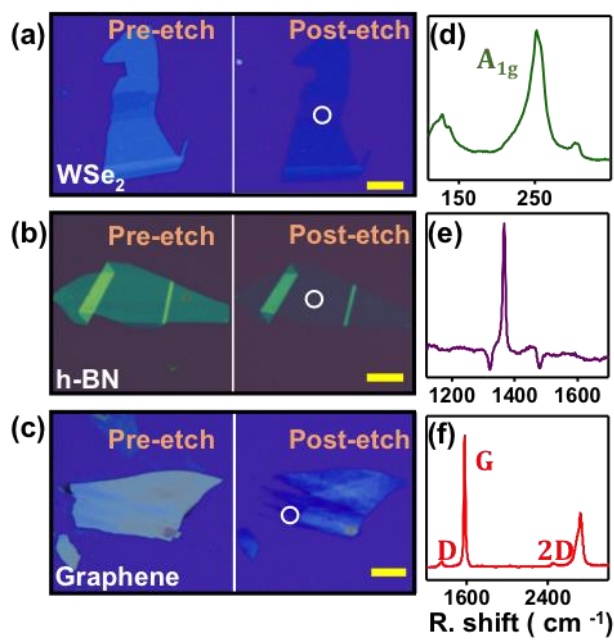

Figure 4: (a), (b) and (c) shows pre-etch and post-etch optical images of $WSe_2$, h-BN and graphene respectively and, (d), (e) and (f) show the respective Raman Spectra, the vertical axis corresponds to the intensity in arbitrary units. The white circles represent the positions from where the Raman spectra were obtained. All the scale bars are 20 µm.

All the structural and morphology analysis suggest that the process is layer-by-layer, preserves the layer-topography and yields a smooth and pristine-like surface. We do not observe any residues, defects or partially etched regions; the post-etch Raman spectra do not show any signatures of oxide residues or defects. The TEM images also confirm that the process has not significantly contributed any point defects or disorders. Sulphur vacancies are known to cause n-type behaviour on pristine exfoliated $MoS_2$[43] and residues of $MoO_3$ is known to cause p-type behaviour.[27] Our transport data exhibit clear n-type behaviour similar to the pristine exfoliated samples. We also note that the threshold voltages for our samples are similar to the pristine samples implying our process has not noticeably altered the defect density. We observe that the process is equally applicable to other vW systems irrespective of their material and electrical properties. In contrast to a defect driven conventional etching, we believe that the process exploits the weak interlayer vW interaction, common to these materials. Plasma exposure weakens vW interaction between the top-layers promoting a residue-free complete removal of the top layer analogous to a peeling action.[36,44] In addition, the in-situ thermal annealing in the reducing FG atmosphere could be helping to preserve the electrical properties.

In this manuscript, we have introduced a novel and simple microwave plasma based top-down layer engineering strategy for a spectrum of vW systems of arbitrary thickness and area. The etching happens by physical removal of the material exploiting the weak inter-layer van der Waals interaction and does not involve any reactive gasses or chemical reactions. The thickness and the etch-rate analysis confirm that the etching is layer-by-layer and possesses a high degree of controllability with relatively short durations compared to some of the other methods. We find that the structural and morphological properties of the material have not been compromised; the devices exhibit electrical behaviour similar to the pristine ones reported in the literature. We have demonstrated device patterning with the aid of conventional lithography techniques and metal masking. Compatibility of microwave process with the standard device fabrication schemes and the capability to etch large area bulk samples down to the few layer regime retaining the material and electrical qualities makes this technique a viable processing tool for the integration of vW materials into the semiconductor industry.

**Acknowledgements**

Authors acknowledge IISER TVM for the infrastructure and experimental facilities, M. M. Shaijumon and P. S. Anil Kumar for a critical reading of the manuscript. MT acknowledges the financial support from DST-SERB extramural program. AV acknowledges INSPIRE and CHS acknowledges CSIR for the fellowship